\documentclass[fleqn,usenatbib]{mnras}

\RequirePackage{snapshot}

\usepackage{newtxtext,newtxmath}

\usepackage[T1]{fontenc}

\DeclareRobustCommand{\VAN}[3]{#2}
\let\VANthebibliography\thebibliography
\def\thebibliography{\DeclareRobustCommand{\VAN}[3]{##3}\VANthebibliography}

\usepackage{graphicx}	
\usepackage{amsmath}	

\newcommand{\vc}[1]{{\mathbf{#1}}}

\title[Precipitation in the magnetized ICM]{Precipitation plausible: magnetized thermal instability in the intracluster medium 
}

\author[B.D. Wibking et al.]{
Benjamin D. Wibking,$^{1}$\thanks{E-mail: wibkingb@msu.edu (BDW)}
G. Mark Voit,$^{1}$
and Brian W. O'Shea$^{1,2,3}$
\\
$^{1}$Department of Physics and Astronomy, Michigan State University, 567 Wilson Road, East Lansing, MI 48824, USA\\
$^{2}$Department of Computational Mathematics, Science, and Engineering, Michigan State University, 428 South Shaw Lane, East Lansing, MI 48824, USA\\
$^{3}$Facility for Rare Isotope Beams, Michigan State University, 640 South Shaw Lane, East Lansing, MI 48824, USA\\
}

\date{Accepted XXX. Received YYY; in original form ZZZ}

\pubyear{2024}

\begin{document}
\label{firstpage}
\pagerange{\pageref{firstpage}--\pageref{lastpage}}
\maketitle

\begin{abstract}
Observations of galaxy-cluster cores reveal that AGN feedback is strongly associated with both a short central cooling time ($t_{\rm c} \lesssim 10^9 \, {\rm yr}$) and accumulations of cold gas ($\lesssim 10^4 \, {\rm K}$). Also, the central ratio of cooling time to freefall time is rarely observed to drop below $t_{\rm c}/t_{\rm ff} \approx 10$, and large accumulations of cold gas are rarely observed in environments with $t_{\rm c} / t_{\rm ff} \gtrsim 30$. Here we show that the critical range --- $10 \lesssim t_{\rm c}/t_{\rm ff} \lesssim 30$ --- plausibly results from magnetized thermal instability. We present numerical simulations of magnetized stratified atmospheres with an initially uniform magnetic field. Thermal instability in an otherwise static atmosphere with $t_{\rm c}/t_{\rm ff} \approx 10$ progresses to nonlinear amplitudes, causing cooler gas to accumulate, as long as the background ratio of thermal pressure to magnetic pressure is $\beta \lesssim 100$. And in atmospheres with $t_{\rm c}/t_{\rm ff} \approx 20$, cooler gas accumulates for $\beta \lesssim 10$. Magnetized atmospheres are therefore much more likely to precipitate than unmagnetized atmospheres with otherwise identical properties. We hypothesize that AGN feedback triggered by accumulations of cold gas prevents $t_{\rm c}/t_{\rm ff}$ from dropping much below 10, because cold gas inevitably precipitates out of magnetized galactic atmospheres with lower ratios, causing $t_{\rm c}/t_{\rm ff}$ to rise.
\end{abstract}




\section{Introduction}

Radiative energy losses from the cores of many galaxy clusters can remove their thermal energy on timescales shorter than a few gigayears. Without a compensating heat source, the central atmosphere of such a cluster would undergo a runaway `cooling flow,' resulting in a star formation rate in the brightest cluster galaxy (BCG) exceeding $\sim 100 \, M_{\odot}$ \citep{Fabian_1984,Fabian_1991}. The observed star formation rates in BCGs are usually only $\sim$1\% to $\sim$10\% of the runaway cooling rate, indicating that pure cooling flows are not operating in the vast majority of galaxy clusters \citep[e.g.,][and references therein]{McDonald_2018}. Instead, outbursts of energy from a central active galactic nucleus (AGN) are thought to compensate for the radiative energy losses \citep[e.g.,][]{Churazov_2001,McNamaraNulsen_2007,McNamaraNulsen_2012}, preventing much of the intracluster gas from reaching the temperatures and densities at which stars form.  

Keeping the time-averaged heating rate similar to the time-averaged cooling rate would seem to require a feedback mechanism that somehow links AGN fueling with cooling of the intracluster medium \citep[e.g.,][]{BinneyTabor_1995,Donahue_2022}. A particularly promising class of feedback models, often called ``precipitation'' or ``chaotic cold accretion'' models, posits that cooling of the hot, volume-filling atmosphere produces a rain of cold clouds that supply fuel to the black hole's accretion engine \citep[e.g.,][]{PizzolatoSoker_2005,McCourt_2012,Sharma_2012,Gaspari_2012ApJ...746...94G,Gaspari_2013,Prasad_2015ApJ...811..108P,Li_2015ApJ...811...73L,Voit_2017}. Those models are considered promising for two reasons: i) the presence of clouds much cooler than the volume-filling atmosphere of a galaxy-cluster core is highly correlated with signatures of strong AGN feedback \citep[e.g.,][]{Cavagnolo_2008}, and ii) there appears to be a nearly universal minimum ratio of cooling time ($t_{\rm c}$) to freefall time ($t_{\rm ff}$) in both galaxy cluster cores and in massive elliptical galaxies. The observed floor is at $t_{\rm c}/t_{\rm ff} \approx 10$ when the local freefall time at radius $r$ in a potential well with gravitational acceleration $g$ is defined to be $t_{\rm ff} = (2r / g)^{1/2}$ and $t_{\rm c}$ is defined to be the thermal energy per unit mass divided by the radiative loss rate per unit mass \citep{Voit_2015Natur.519..203V}.  Also, multiphase gas with a molecular component in the mass range $10^9 \, M_\odot$--$10^{11} \, M_\odot$ is almost always detected in cluster cores with $10 \lesssim t_{\rm c}/t_{\rm ff} \lesssim 30$ \citep[e.g.,][]{Hogan_2017,Pulido_2018ApJ...853..177P}.

However, a simple understanding of why the floor is at $t_{\rm c}/t_{\rm ff} \approx 10$, and not a smaller value, has been hard to come by. Complicated numerical simulations produce precipitation and chaotic cold accretion that apparently self-regulates near that value, at least to within a factor of $\sim 2$ \citep{Gaspari_2012ApJ...746...94G, Meece_2015, Li_2015ApJ...811...73L, Prasad_2015ApJ...811..108P, Tremmel_2019, Nobels_2022}, but more idealized numerical experiments typically do not precipitate at $t_{\rm c}/t_{\rm ff} \approx 10$ \citep[e.g.,][]{McCourt_2012}. 

The simulations performed by \citet{McCourt_2012} have been the archetype. In those simulations, thermal instability led to precipitation for $t_{\rm c}/t_{\rm ff} \lesssim 1$ in atmospheres with volume-averaged heating balancing radiative losses at each altitude. According to a linear stability analysis, perturbation growth should happen on a timescale similar to $t_{\rm c}$, and so overall heating-cooling balance was artificially enforced, so as to prevent the whole atmosphere from cooling and collapsing before precipitation can develop. However, \citet{McCourt_2012} found that precipitation saturates at a density contrast $\delta \rho / \rho \sim ( t_{\rm c} / t_{\rm ff} )^{-1}$ in atmospheres with $t_{\rm c} \gg t_{\rm ff}$ because of non-linear damping, in accordance with analytical expectations \citep[e.g.,][]{Nulsen_1986,Voit_2017}.

Both turbulence \citep{Gaspari_2013} and bulk uplift \citep{McNamara_2016} are capable of promoting thermally unstable cooling and precipitation in atmospheres having a global $t_{\rm c} / t_{\rm ff}$ ratio significantly greater than unity, as long as they produce perturbations in which the \textit{local} ratio is $t_{\rm c} / t_{\rm ff} \lesssim 1$ \citep[see also][]{Voit_2018,Voit_2021}. Motivated by those findings, \citet[][hereafter Paper I]{Wibking_2025} recently performed a suite of simulations designed to clarify the role of turbulence. Global thermal balance was enforced using a mechanism similar to \citet{McCourt_2012}, and several driving modes were employed to drive turbulence having the desired properties. We found that turbulence can indeed raise the critical $t_{\rm c} / t_{\rm ff}$ ratio for non-linear precipitation, but precipitation did not develop in any of our simulations with $t_{\rm c} / t_{\rm ff} \approx 10$. Even in the most favorable case, with vertical driving of turbulence, precipitation failed to appear above $t_{\rm c} / t_{\rm ff} \approx 5$.

We therefore decided to try something else: magnetic fields. \citet{Nulsen_1986} suggested that ``magnetic pinning" might help thermal instability to develop by preventing buoyancy from pulling high-density gas blobs to lower altitudes before they could become full-fledged cold clouds. \citet{Lowenstein_1990} and \citet{Balbus_1991} soon followed up with more complete analyses of what they called \textit{magnetothermal instability}. They showed that even a weak magnetic field qualitatively changes how thermal instability operates because small-scale deformations of a weak magnetic field can counteract the buoyancy effects that would otherwise damp perturbation growth.

The first three-dimensional simulations to produce precipitation in magnetized atmospheres thermally regulated as in \citet{McCourt_2012} were performed by \cite{Ji_2018}. They showed that increasing the magnetic field strength increased the critical $t_{\rm c} / t_{\rm ff}$ ratio at which a thermally regulated atmosphere would precipitate. They also showed that the effects of magnetic fields on precipitation were largely independent of field direction: vertical fields promoted precipitation as readily as horizontal ones. However, they did not explore atmospheres with $t_{\rm c} / t_{\rm ff} \gtrsim 6$ and therefore did not probe the regime of greatest observational interest, with $10 \lesssim t_{\rm c} / t_{\rm ff} \lesssim 30$.

Here, we add magnetic fields to the simulation environment we developed in Paper I and explore the parameter space up to $t_{\rm c} / t_{\rm ff} \approx 30$. We find that magnetic fields can promote precipitation up to $t_{\rm c} / t_{\rm ff} \approx 20$, even when magnetic pressure is relatively insignificant compared to thermal pressure. This finding may explain, at long last, why galactic atmospheres in the range $10 \lesssim t_{\rm c} / t_{\rm ff} \lesssim 30$ tend to be multiphase, with large density and temperature contrasts, and seldom have $t_{\rm c} / t_{\rm ff} \lesssim 10$ (e.g., \citealt{Hogan_2017}).

The rest of this paper is structured as follows: in section \ref{section:methods}, we describe the numerical methods and initial and boundary conditions adopted for our simulations; in section \ref{section:results}, we describe the parameter space of our simulations and the outcomes of thermal instability in our simulations; in section \ref{section:discussion}, we discuss the observational implications of our results and provide suggestions for future work. Section \ref{section:conclusion} summarizes our findings.

\color{black}

\section{Methods}
\label{section:methods}

\subsection{Equations}
The governing equations are those of inviscid magnetohydrodynamics with source terms, written here in units where the magnetic permeability $\mu = 1$:
\begin{align}
    \frac{\partial \rho}{\partial t} + \nabla \cdot (\rho \vc{v})                              & = 0 \, ,       \\
    \frac{\partial (\rho \vc{v})}{\partial t} + \nabla \cdot (\rho \vc{v} \vc{v} - \vc{B} \vc{B} + \mathsf{P}) & = \rho \vc{g} + \rho \vc{a} \, ,  \\
    \frac{\partial E}{\partial t} + \nabla \cdot \left[(E + \mathsf{P})\vc{v} - \vc{B}(\vc{B} \cdot \vc{v})\right]           & = \vc{v} \cdot (\rho \vc{g} + \rho \vc{a}) + \mathcal{H} - \mathcal{C} \, , \\
    \frac{\partial \vc{B}}{\partial t} - \nabla \times (\vc{v} \times \vc{B}) & = 0 \, ,
\end{align}
where $\rho$ is the gas density, $\vc{v}$ is the fluid velocity, $\vc{B}$ is the magnetic field, $E$ is the total energy density (kinetic, thermal, and magnetic),
$\mathsf{P}$ is the total pressure tensor $\mathsf{P} = (P_g + \vc{B}^2 / 2) \, \mathsf{I}$ (with $P_g$ as the gas pressure and $\mathsf{I}$ as the identity tensor), $g$ is the gravitational acceleration, $a$ is the external turbulent acceleration, $\mathcal{H}$ is the heating rate per unit volume, and $\mathcal{C}$ represents cooling per unit volume.

\subsection{Numerical methods}
In order to solve these equations, we use the AthenaPK magnetohydrodynamics (MHD) code, which is based on the \texttt{Parthenon} adaptive mesh refinement (AMR) framework \citep{Grete_2022}. We adopt second-order Runge-Kutta (RK2) time integration with piecewise-linear reconstruction in the primitive variables (density, velocity, and pressure). Unlike in Paper I, we do not use the piecewise-parabolic method (PPM) for reconstruction, because we find that this method is prone to causing numerical instabilities in simulations with strong magnetic fields. In order to suppress the appearance of numerical magnetic monopoles, we use the generalized Lagrange multiplier magnetohydrodynamics (GLM-MHD) method of \cite{Dedner_2002} and \cite{Mignone_2010}.

As in Paper I, we adopt a sub-cell representation of the gas pressure using the method of \cite{Kappeli_2014} to ensure that we hold the correct hydrostatic equilibrium on our finite resolution numerical grid. In order to reduce excess numerical dissipation when in tight hydrostatic equilibrium, we include the low-Mach correction of \cite{Minoshima_2021} in the implementation of the Harten-Lax-van Leer with Discontinuities (HLLD; \citealt{Miyoshi_2005}) Riemann solver.

The cooling term is included in the simulation via first-order operator splitting, where the operator split step evolves the internal energy in each cell using an adaptive-timestep second-order Runge-Kutta integrator. The form of the cooling term is
\begin{align}
    \mathcal{C} = \rho^2 \Lambda
\end{align}
where $\Lambda$ is assumed to be constant independent of temperature. We choose this cooling function for simplicity, since the linear analysis of thermal instability indicates only an order-unity dependence on the temperature exponent (e.g., \citealt{Donahue_2022}). The overall density normalization of our simulations is then adjusted in order to produce simulations with differing cooling times.

The heating term is calculated using the `thermostat' mechanism originally described in Paper I. To briefly summarize, we add a term proportional to the difference between the current volume-average temperature at a given height $\langle T \rangle_{z}$ and the initial temperature $T_0$ in order to both prevent a global cooling flow and enforce a nearly-constant vertical entropy profile (averaged over a cooling time), while simultaneously allowing for local thermal instabilities:
\begin{align}
    \mathcal{H} = - \frac{K_p}{t_{\text{cool}}(z)} \, \rho c_v \, \langle \Delta T \rangle_z \,  ,
\end{align}
where $K_p$ is a dimensionless constant of proportionality, $t_{\text{cool}}(z)$ is the mean cooling time at height $z$, $c_v$ is the specific heat capacity of the gas, and $\langle \Delta T \rangle_{z} = \langle T \rangle_{z} - T_0$ is the is volume-averaged mean difference between the local temperature and the background temperature profile at height $z$. We note that this `heating' term can be negative when the average temperature at a given height is greater than the initial temperature of the simulation. This term drives the volume-average mean temperature at all heights back to the initial temperature of the simulation on a timescale comparable to the local cooling time $t_{\text{cool}}(z)$. As a result, the volume-averaged vertical entropy profile is relatively constant (within 10--20 per cent) as each simulation evolves over many cooling times. In Paper I, it was shown that this mechanism produces qualitatively different outcomes for thermal instability in CGM environments as compared to earlier simulations in the literature without it (e.g., \citealt{Sharma_2012,Gaspari_2013}), since simulations lacking such a mechanism appear to exhibit a secular evolution of their background entropy profile  via turbulent diffusion toward an isentropic state.

Finally, we set the gravitational acceleration to a nearly constant absolute value $g_0$, except near the midplane of the simulation. Since the sign of the gravitational acceleration is negative in the upper part of the box and positive in the lower part, we have reduced its magnitude near the midplane in order to smoothly transition between $-g_0$ and $+g_0$. It is unchanged from the gravitational acceleration term used in Paper I. We note that the heating and cooling terms are both disabled within the region where the gravitational acceleration smoothly transitions to zero and changes sign ($|z| \lesssim 5$ kpc; denoted the `exclusion zone') in order to prevent excessive production of cold gas, since the lack of stratification could allow thermal instability to grow in an uncontrolled way that is not representative of the typical value of $t_{\text{c}} / t_{\text{ff}}$ at higher altitudes.

In contrast with Paper I, there is no external acceleration field used to drive turbulence in the simulations used in this work. All of our simulations develop turbulence but it is generated purely by fluid instabilities. This result is consistent with the seminal simulations of thermal instability in galaxy clusters of \cite{McCourt_2012} and \cite{Sharma_2012}. Future work will consider the extension to a larger parameter space that includes external turbulent driving (as in Paper I) in order to represent motions induced by AGN jets and other large-scale sources of turbulence.

\subsection{Parameter space}
We construct the parameter space for our suite of simulations by using a quasi-random sampling of a two dimensional parameter space. The two parameters considered are the ratio of cooling time to free-fall time $t_c/t_{\text{ff}}$ and the plasma beta $\beta$ at an altitude of $10 \, {\rm kpc}$, where the plasma beta is defined as $\beta \equiv P_{\text{gas}}/P_{\text{mag}}$, where $P_{\text{gas}} = n k_B T$ and $P_{\text{mag}} = B^2 / 2$\footnote{In Heaviside-Lorenz units, $B$ differs from its value in CGS-Gaussian units by a factor of $1/\sqrt{4\pi}$.}. We consider $t_c/t_{\text{ff}}$ in log space from 1 to 30, and $\beta$ in log space from 1 to 1000. Samples are chosen with a uniform density in this two-dimensional parameter space using the \cite{Halton_1960} sequence. We choose to use a quasi-random sequence instead of uniform random sampling in each parameter in order to more efficiently sample the parameter space by avoiding clusters of points in parameter space that occur by chance (e.g., \citealt{Berblinger_1991}).

The initial conditions of our simulations are identical to those of Paper I, except for the inclusion of a horizontal magnetic field. To summarize, we initialize the gas in a state of hydrostatic equilibrium assuming a constant temperature ($T_0 = 10^7$ K), with per cent-level density perturbations in order to seed instabilities. Since the gravitational acceleration is nearly constant, the isothermal atmosphere has exponential pressure and density profiles, with a scale height of 50~kpc. The box height is twice the pressure scale height (100~kpc) in each direction away from the midplane, resulting in a background value of $t_{\text{c}}/t_{\text{ff}}$ that varies by less than $\approx 20$ per cent between the altitudes of 5 kpc and 50 kpc (see Paper I). This setup allows us to resolve the local pressure scale height with many cells everywhere while also maintaining a nearly constant value of $t_{\text{c}}/t_{\text{ff}}$. Since the background magnetic field has no gradient, the hydrostatic pressure and density profiles are unchanged. While we consider only horizontal fields, \cite{Ji_2018} also explored vertical fields and found that they have very similar effects on the density contrast $\delta \rho / \rho$ at which thermal instability saturates. We adjust the plasma beta parameter by varying the strength of the spatially-constant horizontal magnetic field in the initial conditions, and the cooling time to freefall time ratio $t_c/t_{\text{ff}}$ is adjusted by varying the normalization of the background density profile.

\section{Results}
\label{section:results}

\begin{figure*}
    \includegraphics[width=\textwidth]{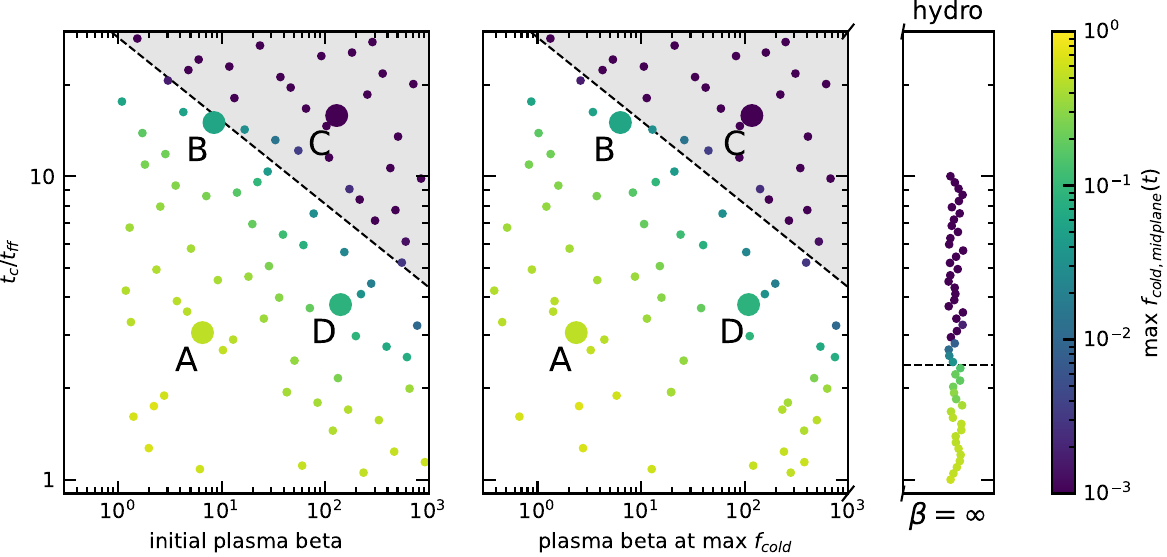}
    \caption{Dependence of cold gas accumulation on the mean value of $\beta$ at 10~kpc and the approximate $t_{\text{c}}/t_{\text{ff}}$ ratio in the 5 kpc to 50 kpc altitude range. The left panel shows the initial $\beta$ and the central panel shows the mean $\beta$ at the moment of maximum cold-gas accumulation. The corresponding hydrodynamic simulations without magnetic fields ($\beta = \infty$, right panel) are shown as points in a nearly vertical line (with horizontal jitter added to help separate the individual points). A color bar shows how cold-gas accumulation is represented. Note that the simulations with magnetic fields ($\beta \lesssim 10^3$) have non-trivial amounts of cold gas at higher values of $t_{\text{c}}/t_{\text{ff}}$ than in the hydrodynamic simulations. Points labeled with letters indicate simulations highlighted in Figures \ref{fig:temperature_pdf} and \ref{fig:volume_renders}.}
    \label{fig:beta_vs_tctff_nodriving}
\end{figure*}

Figure \ref{fig:beta_vs_tctff_nodriving} shows 100 simulations randomly distributed in a two-dimensional parameter space with  $\beta$ (evaluated at an altitude of 10~kpc) along the horizontal axis and $t_c/t_{\text{ff}}$ (evaluated at the pressure scale height of 50~kpc) along the vertical axis. The parameter $\beta$ is either the value in the initial conditions (shown in the left panel) or the volume-averaged value of $\beta$ at $10$~kpc at the moment of maximum cold-gas accumulation (shown in the middle panel). Each point represents an individual simulation, color-coded by the maximum mass fraction $f_{\text{cold}}$ of cold gas ($T < 2 \times 10^6$ K) near the midplane (in the volume within $|z| < 10$ kpc), among all of the outputs of a given simulation. We find that the mean $\beta$ at 10~kpc does not significantly evolve during our simulations.

\begin{figure}
    \includegraphics[width=\columnwidth]{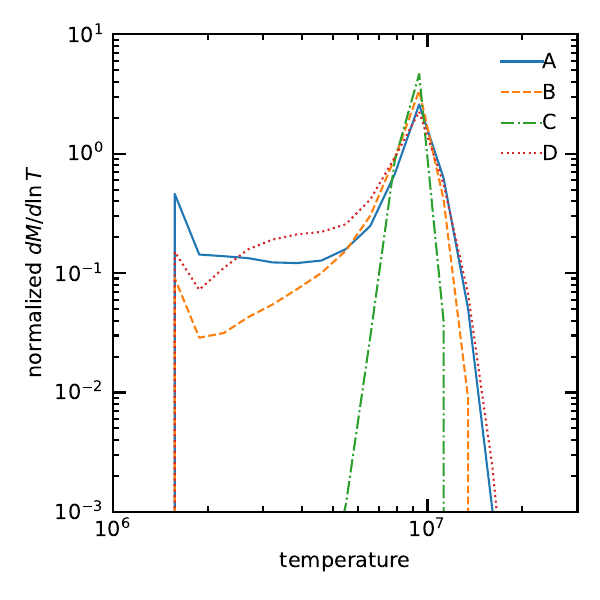}
    \caption{Mass-weighted probability distribution functions (PDFs) for temperature at the moment of maximum cold-gas accumulation. Simulation A (solid line, $\beta \approx 10$, $t_{\rm c}/t_{\rm ff} \approx 3$) has the broadest PDF and the greatest accumulation of cold gas. Simulation B (dashed line, $\beta \approx 10$, $t_{\rm c}/t_{\rm ff} \approx 15$) has a narrower PDF but still produces cold gas. Simulation C (dot-dashed line, $\beta \approx 100$, $t_{\rm c}/t_{\rm ff} \approx 15$) has the narrowest PDF and essentially no cold gas. Simulation D (dotted line, $\beta \approx 100$, $t_{\rm c}/t_{\rm ff} \approx 4$) has the broadest PDF and cold-gas accumulation similar to simulation B.}
    \label{fig:temperature_pdf}
\end{figure}

\begin{figure*}
    \includegraphics[width=\textwidth]{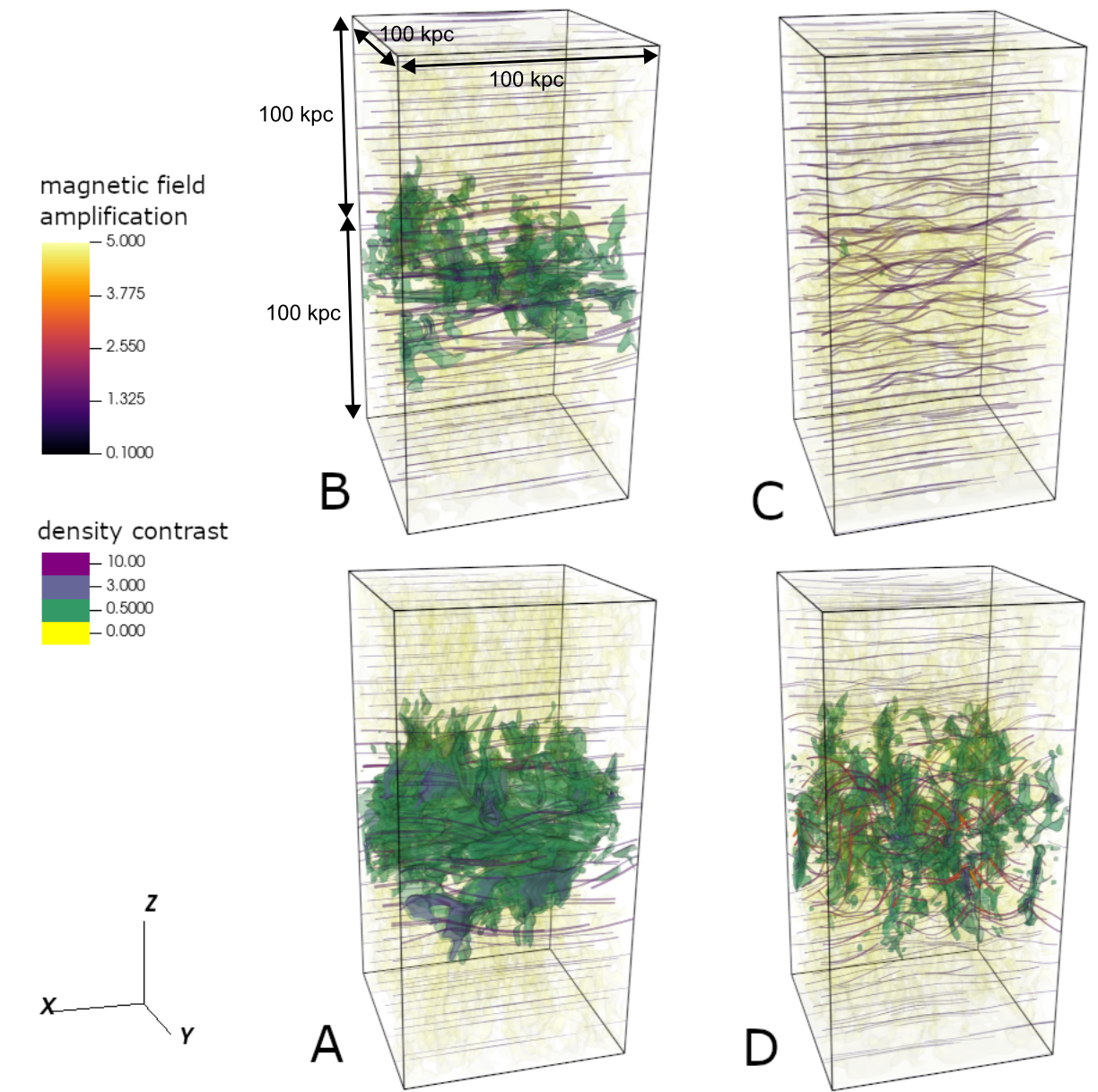}
    \caption{Volume renderings of simulations B and C (first row) and A and D (second row). The colorbar associated with the magnetic field lines indicates the magnetic field amplification with respect to the initial conditions. The color-coded surfaces indicate contours of constant density contrast with respect to the local mean density at a given height. Each simulation volume is 100~kpc in width and extends vertically 100~kpc from the midplane, in both directions.}
    \label{fig:volume_renders}
\end{figure*}

Figure \ref{fig:temperature_pdf} shows the probability density function of gas temperature, weighted by mass, for four simulations drawn from representative regions of the Figure \ref{fig:beta_vs_tctff_nodriving} parameter space. In particular, Simulation A is representative of the quadrant with low $\beta$ and low $t_{\text{c}}/t_{\text{ff}}$. Simulation B is representative of the quadrant with low $\beta$ and high $t_{\text{c}}/t_{\text{ff}}$. Simulation C is representative of the quadrant with high $\beta$ and high $t_{\text{c}}/t_{\text{ff}}$. Simulation D is representative of the quadrant with high $\beta$ and low $t_{\text{c}}/t_{\text{ff}}$.

Simulations with low $t_{\text{c}}/t_{\text{ff}}$ (represented by A and D) all develop significant amounts of cold gas ($f_{\text{cold}} \gtrsim 0.1$) near the midplane, and so do the simulations with low $\beta$ (represented by A and B). Figure \ref{fig:temperature_pdf} shows that simulations A, B, and D also have broader temperature distributions than simulation C around the peak at $\approx 10^7 \, {\rm K}$, along with asymmetric tails extending toward lower temperatures. We will refer to simulations like A, B, and D as ``multiphase'' and simulations like C as ``single-phase.''

The dividing line in parameter space between single-phase and multiphase simulations is approximately given by the power law:
\begin{align}
    \log_{10} (t_{\text{c}}/t_{\text{ff}}) &= -0.3 \, \log_{10} \beta + 1.5
\end{align}

\noindent We obtain the slope and intercept by fitting a linear support vector machine (SVM) in the log-log parameter space shown in Figure \ref{fig:beta_vs_tctff_nodriving}. We label the simulation points where $f_{\text{cold}} < 0.1$ as single phase and points where $f_{\text{cold}} > 0.1$ as multiphase. Then the linear kernel SVM produces the line that `best' separates these two classifications in the parameter space of $\log_{10} \beta$ and $\log_{10} t_{\text{c}}/t_{\text{ff}}$. For a description of the method, see Appendix \ref{appendix:svm}.

However, below $\beta \lesssim 10$, it appears that the critical dividing line between multiphase and single-phase simulation outcomes may becloser to a horizontal line, i.e., simulations have multiphase outcomes for $t_{\text{c}}/t_{\text{ff}} \lesssim 20$. Due to the sparseness of the sampling in this region of parameter space, we do not attempt to separately fit the $\beta \lesssim 10$ parameter space using the points shown in Figure \ref{fig:beta_vs_tctff_nodriving}.

\section{Discussion}
\label{section:discussion}
\subsection{Observational significance}
The most significant result of this work is illustrated by the outcome of simulation B: multiphase gas accumulates in an environment with a significantly larger value of $t_{\text{c}}/t_{\text{ff}}$ than in any of Paper I's simulations without magnetic fields. The setup is otherwise identical. Also, no other similar purely hydrodynamical simulation of a stratified galactic atmosphere has produced multiphase gas, as far as we know.\footnote{Some of the simulations by \citet{Meece_2015} produced multiphase gas in atmospheres with similarly large values of $t_{\rm c}/t_{\rm ff}$ away from the midplane, but that happened because cooling and heating were allowed to operate near the midplane, where gravity vanished and $t_{\rm c}/t_{\rm ff}$ became small \citep[see][for a discussion]{Voit_2017}.} Paper I showed that idealized hydrodynamic simulations of precipitation did not produce multiphase gas for $t_{\text{c}}/t_{\text{ff}} \gtrsim 5$, even with significant extrinsic turbulent driving. Furthermore, multiphase gas was not produced without external driving for any values of $t_{\text{c}}/t_{\text{ff}} \gtrsim 3$. With these null results in mind, it is highly significant that multiphase gas is produced in simulations B and D, without any external driving.

As mentioned in the introduction, observations of galaxy clusters find that the presence of multiphase gas is correlated with a critical threshold value of $t_{\text{c}}/t_{\text{ff}}$ between 10 and 30 (e.g., \citealt{Cavagnolo_2009}, \citealt{Hogan_2017}, \citealt{Donahue_2022}). Figure \ref{fig:beta_vs_tctff_nodriving} shows that multiphase gas is produced in our idealized simulations with magnetic fields for values of $t_{\text{c}}/t_{\text{ff}}$ as high as 20. While not a `smoking gun,' this value lies tantalizingly close to the center of the observational range of cooling-to-free-fall times observed to be correlated with the presence of multiphase gas, suggesting that precipitation assisted by magnetic fields may play a critical role in determining the presence of multiphase gas in galaxy clusters.

Likewise, there is a long history of measurements of the magnetic fields of galaxy clusters (see \citealt{Carilli_2002} and references therein). The observational constraints on the magnetic field strength of the intracluster medium (ICM) within galaxy clusters are uncertain to within about one order of magnitude due to fundamental systematic uncertainties in using Faraday rotation to estimate the field strength \citep{Johnson_2020}. Some uncertainties can be mitigated by measuring the depolarization of background sources, finding a central magnetic field strength between 5 and 10 $\mu G$ \citep{Osinga_2022}. Combining rotation measure and depolarization information for \emph{Planck}-selected massive clusters, \cite{Osinga_2025} obtains a mean line-of-sight magnetic field strength of
\begin{align}
B_0 \approx 2 \, \left( \frac{n_e}{10^{-3} \, \text{cm}^{-3}} \right) \, \left( \frac{l}{10 \, \text{kpc}} \right)^{‑1/2} \, \left( \frac{L}{1 \, \text{Mpc}} \right)^{‑1/2}\, \mu \text{G}
\end{align}
under the assumption of a constant, uniform field strength and electron number density, where $n_e$ is the electron number density, $l$ is the scale on which the direction of the magnetic field fluctuates, and $L$ is the line-of-sight integration length.\footnote{Estimates of the mean field strength in galaxy groups obtain similar values (e.g., \citealt{Anderson_2024}).} A more complex model that includes radial variation of the magnetic field strength and electron density yields best fit magnetic field strengths of either $B_0 \approx 5 \, \mu \text{G}$ or $10 \, \mu \text{G}$, depending on whether depolarization or rotation measure observations are preferred \citep{Osinga_2025}. Converting these estimates of the magnetic field strength into estimates of plasma beta requires additional assumptions about the galaxy cluster pressure profile. However, a simple estimate can be made by assuming a constant density equal to the value used for the simplest Faraday rotation-based field estimate and a temperature close to the virial temperature. Then we have:
\begin{align}
\beta \sim 50 \, \left( \frac{n_e}{10^{-3} \text{cm}^{-3}} \right) \, \left( \frac{T}{10^8 \text{K}} \right) \, \left( \frac{B_0}{1 \, \mu \text{G}} \right)^{-2} \, .
\end{align}
Since this value of plasma beta lies well within the regime where magnetic fields make a significant difference for thermal instability (Figure \ref{fig:beta_vs_tctff_nodriving}), we conclude that our results are directly applicable to the intracluster medium.

\subsection{Previous work}
Our work is closely related to that of \cite{Ji_2018}, who used a simulation setup very similar to that used here, except that they did not employ a `thermostat' mechanism like ours in order to control the background entropy gradient. Our results qualitatively agree, as both efforts find that the presence of magnetic fields promotes development of nonlinear density contrasts, even for plasma beta $\beta \gg 1$. \cite{Ji_2018} explored a range of plasma beta values $\beta$ between about $0.4$ and $300$, and $t_{\text{c}}/t_{\text{ff}}$ values between about $0.3$ and $6$. However, they did not sample the range of greatest interest to observers of galaxy cluster cores: $10 \lesssim t_{\text{c}}/t_{\text{ff}} \lesssim 30$. Our work has extended the $t_{\text{c}}/t_{\text{ff}}$ dimension in order to sample that range.

\subsection{Thermal conduction}
Neither our simulations nor those of \cite{Ji_2018} include thermal conduction, which could potentially suppress development of large density contrasts and fundamentally changes the qualitative nature of magnetothermal instability if it is strong \citep[e.g.,][]{Balbus_1991}. The usual approach to assessing the importance of thermal conduction is to estimate a length scale called the \textit{Field length} (e.g., \citealt{McKee_1977}):
\begin{align}
    \lambda_{\rm F} \equiv \sqrt{ \frac {\kappa T} {n^2 \Lambda_e} } 
    \; \; ,
\end{align}
where $\kappa \approx ( 5 \times 10^{-7} \, {\rm erg \, s^{-1} \, cm^{-1} \, K^{-7/2}}) \, T^{5/2}$ is the Spitzer conduction coefficient (assuming a constant Coulomb logarithm), and $\Lambda_e \equiv (\rho/n_e)^2 \Lambda$ is a cooling function defined with respect to electron density. If conduction proceeds at the full Spitzer rate, then it inhibits thermal instability along magnetic field lines on scales smaller than $\lambda_F$. The Field length is
\begin{align}
    \begin{split}
    \lambda_{\rm F} 
        \approx \, 2.5 \, \text{kpc} \, 
            &\left( \frac{\Lambda_e}{3 \times 10^{-23} \, \text{erg} \,  \text{cm}^3 \, \text{s}^{-1}} \right)^{-1/2} \, \times \\ 
            &\left( \frac{T}{10^{7} \, \text{K}} \right)^{7/4} \left( \frac{n_e}{0.03 \, \text{cm}^{-3}} \right)^{-1} \, f_{\text{Spitzer}}^{1/2} ,
    \end{split}
\end{align}
where $f_{\text{Spitzer}}$ is a suppression factor for conduction relative to the Spitzer conductivity.
Cluster cores with multiphase gas typically have $kTn_e^{-2/3} \sim 10 \, {\rm keV \, cm^2}$ at radii $\sim 10$~kpc \citep{Donahue_2006ApJ...643..730D}, meaning that thermal instability can proceed on scales similar to the atmospheric scale height, but formation of cold clouds on much smaller scales requires significant suppression of Spitzer-like conduction. Recent experimental results suggest that conduction is suppressed relative to the Spitzer value in collisionless high-beta turbulent plasmas by a factor of at least $\sim 100$ \citep{Meinecke_2022}, i.e., $f_{\text{Spitzer}} \sim 0.01$, corresponding to a reduction of the Field length by at least one order of magnitude.

\subsection{Future work}
One obvious extension of the present work is to include external sources of turbulent driving, e.g., as considered in \cite{Wibking_2025}. Since both external turbulent driving and magnetic fields separately facilitate the formation of cold gas beyond their initial cooling-to-free-fall ratio limits, we suspect that the combination of turbulence and magnetic fields will have an even more significant effect in increasing the cold gas formation in simulations and raising the critical value of $t_{\text{c}}/t_{\text{ff}}$ for a large range of values of plasma beta. Due to the different effects of the driving mechanism, driving amplitude, plasma beta, and ratio of cooling-to-free-fall time, the  parameter space of interest is large. A set of simulations to explore this effect is in progress and will be discussed in a future paper.

Another extension of the present work is to consider variations of the magnetic field geometry in the initial conditions. \cite{Ji_2018} considered vertical fields, finding that the nonlinear outcome of thermal instability in simulations with vertical fields saturates at a similar $\delta \rho / \rho$ scale but with a different morphology than in their simulations with initially-horizontal fields. We therefore anticipate that the saturation results we have presented here do not depend critically on the magnetic field direction. However, fields with no net flux (i.e., that are not attached to the boundary) may yield a different outcome, since the field dynamics are then no longer directly tied to the boundary conditions of the simulation. In particular, it may be interesting to explore the effects of tangled fields (or field loops) with varying coherence lengths. Magnetic fields with coherence lengths much less and much greater than the scale height of the atmosphere, respectively, are limiting cases of this geometry, and we plan to simulate these limits in future work.

\section{Conclusion}
\label{section:conclusion}

In our idealized simulations of stratified and thermally balanced atmospheres, we find that magnetic fields significantly assist precipitation of clouds through thermal instability, including in regions of parameter space that were previously unexplored. We summarize our results as follows:
\begin{itemize}
    \item For strong magnetic fields ($\beta \sim 1$), cold clouds form for cooling-to-free-fall ratio values of  $t_{\text{c}}/t_{\text{ff}} \lesssim 20$.
    \item For weak fields ($\beta \gg 1$), we find that the presence of magnetic fields significantly increases the amount of cold gas compared to the results of our otherwise identical purely hydrodynamic simulations, even the ones with driven turbulence, up to $\beta \sim 10^{3}$.
    \item For $\beta \sim 1$, the transition in parameter space between multiphase and non-multiphase simulations is very nearly a horizontal line in Figure \ref{fig:beta_vs_tctff_nodriving}, and is around $t_{\text{c}}/t_{\text{ff}} \sim 20$.
    \item For $\beta \gg 1$, the critical value of $t_{\text{c}}/t_{\text{ff}}$ for multiphase gas formation declines with increasing $\beta$ and is approximately a power law, up to $\beta \sim 10^{3}$, at which point the MHD and hydrodynamic simulation outcomes are approximately the same.
\end{itemize}

Our results showing that precipitation in magnetized atmospheres with $t_{\text{c}}/t_{\text{ff}} \approx 20$ and $\beta \lesssim 10$ are tantalizing, in that they are close to the center of the observationally interesting parameter regime ($10 \lesssim t_{\text{c}}/t_{\text{ff}} \lesssim 30$). While the magnetic field strength in galaxy clusters is uncertain, the volume-weighted plasma beta is likely in the range $10 \lesssim \beta \lesssim 100$ (see discussion). Thus, magnetically assisted precipitation may be important for feedback regulation of galaxy-cluster cores if either i) there are large localized fluctuations in the plasma beta and thermal instability produces significant cold gas in those regions, or ii) the addition of external sources of turbulence allows stratified thermal instability to operate in the moderate $\beta$ regime of the bulk intracluster medium (ICM). Precipitation would then be responsible for a floor at $t_{\text{c}}/t_{\text{ff}} \approx 10$ in the ambient medium, which would otherwise be too unstable to cold-cloud formation. We plan to explore these possibilities in future work.

\section*{Acknowledgements}
We thank Philip Grete and Forrest Glines for technical assistance with the use of the AthenaPK MHD code.  The authors acknowledge the support of U.S. National Science Foundation (NSF) grant \#AAG-2106575, which is the source of support for BDW.

BDW thanks Michael Zingale for suggesting the use of well-balanced reconstruction methods and reflecting vertical boundary conditions for hydrodynamics simulations near hydrostatic equilibrium.

BWO acknowledges support from NSF grant \#AAG-1908109, National Aeronautics and Space Administration (NASA) Astrophysics Theory Program (ATP) grants NNX15AP39G and 80NSSC18K1105, and NASA Theoretical and Computational Astrophysics Networks (TCAN) grant 80NSSC21K1053. 

The simulations were carried out in part using ACCESS resources using allocation MCA08X028 (TG-AST090040; PI: Brian O'Shea). This work was also supported in part through computational resources and services provided by the Institute for Cyber-Enabled Research at Michigan State University.

This research has made use of NASA's Astrophysics Data System.

\emph{Software:} \texttt{Chaospy} \citep{Feinberg_2015}, \texttt{Jupyter} \citep{2007CSE.....9c..21P, kluyver2016jupyter}, \texttt{matplotlib} \citep{Hunter_2007}, \texttt{NumPy} \citep{Harris_2020},  \texttt{Parthenon} \citep{Grete_2022}, \texttt{scikit-learn} \citep{scikit-learn, sklearn_api, scikit-learn_14627164}, \texttt{scipy} \citep{2020SciPy-NMeth, scipy_15484555}, \texttt{tqdm} \citep{tqdm_14231923}, \texttt{VisIt}  \citep{Childs_2012}, \texttt{yt} \citep{2011ApJS..192....9T}.\footnote{Software citation information aggregated using \texttt{\href{https://www.tomwagg.com/software-citation-station/}{The Software Citation Station}} \citep{software-citation-station-paper, software-citation-station-zenodo}.}

\section*{Data Availability}

The simulation data underlying this article are available upon request from the authors.



\bibliographystyle{mnras}
\bibliography{precipitator}



\appendix
\section{Linear support vector machine with two classes}
\label{appendix:svm}
For two classes of points that can be perfectly separated by a line, the line determined by the linear support vector machine (SVM) is the `best' one in the sense that it is the unique line separating the two classes of points that maximizes the sum of the distances between the line and the closest point in each class. Since real data has points that cannot be perfectly separated in this way, the definition of the SVM involves minimizing a functional that includes a penalty term proportional to the square of the weight vector:
\begin{align}
\min_{w,b} \, \alpha \, \frac{1}{2} w^T w + \sum_{i=1}^{N} \max[0, 1 - y_i(w^T x_i + b)]
\end{align}
where $x_i$ are the input points (as $D$-dimensional vectors), $y_i$ are the classifications of the points (as a 2-dimensional vector), $w$ is the $D$-dimensional weight vector and $b$ is proportional to the intercept of the separating line. In our case, $D=2$. We use the implementation of \texttt{scikit-learn} to compute $w$ and $b$ \citep{scikit-learn, sklearn_api, scikit-learn_14627164}. The slope of the separating line is $-w_0/w_1$ and the intercept of the line is $-b/w_1$, where $w_0$ and $w_1$ are the two components of $w$.

The strength $\alpha$ of the penalty term is a hyperparameter and must be determined by other means. We determine the strength $\alpha$ of the penalty term is  by cross-validation (using five folds of the data) with the \cite{Jaccard_1912} similarity index. This sets $\alpha$ by maximizing the predictive classification ability of the SVM for data that is left out of the training set.


\bsp	
\label{lastpage}
\end{document}